\documentclass[doublecol]{epl2}
\usepackage{amsmath,amssymb,graphicx}
\newcommand{\D}{{\mathrm{d}}}

\title{Reciprocal Relations Between Kinetic Curves}
\author{G.S. Yablonsky\thanks{Long Term Structural Methusalem Funding
by the Flemish Government - grant number BOF09/01M00409}\inst{1}
\and A.N. Gorban\thanks{Support from the University of Leicester and
Isaac Newton Institute  for Mathematical Sciences}\inst{2}
 \and D. Constales\thanks{Support from BOF/GOA
01GA0405 of Ghent University}\inst{3}
 \and V.V. Galvita \inst{4}
 \and G.B. Marin\inst{({\rm a})4}}

 \shortauthor{G.S. Yablonsky \etal}

\institute{
  \inst{1} {Parks College, Department of Chemistry, Saint
Louis University - Saint Louis, MO 63103,
USA}\\
  \inst{2} {Department of Mathematics, University of
Leicester - Leicester LE1 7RH, UK}\\
  \inst{3} {Department of
Mathematical Analysis, Ghent University - Galglaan 2, B-9000
Gent, Belgium}\\
  \inst{4} {Laboratory for Chemical Technology, Ghent
  University - Krijgslaan 281 (S5), B-9000 Gent, Belgium}
  }


\abstract{We study coupled irreversible processes. For linear
or linearized kinetics with microreversibility,  $\dot{x}=Kx$,
the kinetic operator $K$ is symmetric in the entropic inner
product. This form of Onsager's reciprocal relations  implies
that the shift in time, $\exp (Kt)$, is also a symmetric
operator. This generates the reciprocity relations between the
kinetic curves. For example, for the Master equation, if we
start the process from the $i$th pure state and measure the
probability $p_j(t)$ of the $j$th state ($j\neq i$), and,
similarly, measure $p_i(t)$ for the process, which starts at
the $j$th pure state, then the ratio of these two probabilities
$p_j(t)/p_i(t)$ is constant in time and coincides with the
ratio of the equilibrium probabilities. We study similar and
more general reciprocal relations between the kinetic curves.
The experimental evidence provided as an example is from the
reversible water gas shift reaction over iron oxide catalyst.
The experimental data are obtained using Temporal Analysis of
Products (TAP) pulse-response studies. These offer excellent
confirmation within the experimental error.}

\begin{document}

\maketitle

\section{Introduction}
\label{sec1}
\subsection{A bit of history}

In 1931, L.~Onsager \cite{Onsager1,Onsager2} gave the
backgrounds and generalizations to the reciprocal relations
introduced in 19th century by Lord Kelvin and H.~v.~Helmholtz.
In his seminal papers, L.~Onsager mentioned also the close
connection between these relations and {\em detailed balancing}
of elementary processes: at equilibrium, each elementary
transaction should be equilibrated by its inverse transaction.
This principle of detailed balance was known long before for
the Boltzmann equation \cite{Boltzmann}. A.~Einstein used this
principle for the linear kinetics of emission and absorption of
radiation \cite{Einstein1916}. In 1901, R. Wegscheider
published an analysis of detailed balance for chemical kinetics
\cite{Wegscheider1901}.

The connections between the detailed balancing and Onsager's
reciprocal relations were clarified in detail by
N.~G.~v.~Kampen \cite{vanKampen1973}. They were also extended
for various types of coordinate transformations which may
include time derivatives and integration in time
\cite{Stockel}. Recently, \cite{Astumian2008}, the reciprocal
relations were derived for nonlinear coupled transport
processes between reservoirs coupled at mesoscopic contact
points. Now, an elegant geometric framework is elaborated for
Onsager's relations and their generalizations
\cite{Grmela2002}.

Onsager's relations are widely used for extraction of kinetic
information about reciprocal processes from experiments and for
the validation of such information (see, for example,
\cite{Ozera2005}): one can measure how  process A affects
process B and extract the reciprocal information, how B affects
A.

The reciprocal relations were tested experimentally for many
systems. In 1960, D.G.~Miller wrote a remarkable review on
experimental verification of the Onsager reciprocal relations
which is often referred to even now \cite{Miller1960}.
Analyzing many different cases of irreversible phenomena
(thermoelectricity, electrokinetics, isothermal diffusion,
etc), Miller found that these reciprocal relations are valid.
However, regarding the chemical reactions, Miller's point was :
``The experimental studies of this phenomenon ... have been
inconclusive, and the question is still open from an
experimental point."

According to Onsager's work \cite{Onsager1}, the fluxes in chemical
kinetics are time derivatives of the concentrations and potentials
are expressed through the chemical potentials. The fluxes (near
equilibrium) are linear functions of potentials and the reciprocal
relations state that the coefficient matrix of these functions is
symmetric. It is impossible to measure these coefficients directly.
To find them one has to solve the inverse problem of chemical
kinetics. This problem is often ill-posed.

Such a difficulty, appearance of ill-posed problems in the
verification of the reciprocal relations, is typical because
these relations connect the kinetic coefficients. Sometimes it
is possible to find them directly in separate experiments but
if it is impossible then the inverse problem arises with all
the typical difficulties.

In our work we, in particular, demonstrate how it is possible
to verify the reciprocal relations without the differentiation
of the empiric kinetic curves and  solving the inverse
problems, and present the experimental results which
demonstrate these relations for one reaction kinetic system.
For this purpose, we have to formulate the reciprocal relations
directly between the measurable quantities.

These reciprocal relations between kinetic curves use the
symmetry of the propagator in the special entropic inner
product.  A dual experiment is defined for each ideal kinetic
experiment. For this dual experiment, both {the initial data
and the observables are different (they exchange their
positions), but the results of the measurement is essentially
the same function of time.}

\subsection{The structure of the paper}

We start from the classical Onsager relations and reformulate
them as conditions on the kinetic operator $K$ for linear or
linearized kinetic equations $\dot{x}=Kx$. This operator should
be symmetric in the entropic inner product, whereas the matrix
$L$ that transforms forces into fluxes should be symmetric in
the standard inner product, i.e. $L_{ij}=L_{ji}$. The form of
reciprocal relations with special inner product is well known
in chemical and Boltzmann kinetics
\cite{Yablonskiiatal1991,GorKar}. They are usually proved
directly from the detailed balance conditions. Such relations
are also universal just as the classical relations are.

Real functions of symmetric operators are also symmetric. In
particular, the propagator $\exp(Kt)$ is symmetric. Therefore,
we can formulate the reciprocal relation between kinetic
curves. These relations do not include fluxes and time
derivatives, hence, they are more robust. {We formulate them as
the symmetry relations between the observables and initial data
(the {\em observables-initial data symmetry}).}

A particular case of this symmetry for a network of
monomolecular chemical reactions or for the Master equation,
which describe systems with detailed balance, seems rather
unexpected. Let us consider two situations for a linear
reaction network.
\begin{enumerate}
\item{The process starts at the state ``everything is in
    $A_q$", and we measure the concentration of $A_r$. The
    result is $c^a_r(t)$ (``how much $A_r$ is produced from
    the initial $A_q$").}
\item{The process starts at the state ``everything is in
    $A_r$", and we measure the concentration of $A_q$. The
    result is $c^b_q(t)$ (``how much $A_q$ is produced from
    the initial $A_r$") (the {\it dual experiment}).}
\end{enumerate}
The results of the dual experiments are connected by the
identity
$$\frac{c^a_r(t)}{c^{\rm eq}_r}\equiv \frac{c^b_q(t)}{c^{\rm
eq}_q} \, ,$$ where $c$ are concentrations and $c^{\rm eq}$
are equilibrium concentrations.

{The symmetry with respect to the observables-initial data
exchange} gives the general rule for production of the
reciprocal relations between kinetic curves.

Many real processes in chemical engineering and biochemistry
include irreversible reactions, i.e. the reactions with a
negligible (zero) rate of the reverse reaction. For these
processes, the micro-reversibility conditions and the
backgrounds of classical Onsager relations are not applicable
directly. Nevertheless, they may be considered as limits of
systems with micro-reversibility when some of the rate
constants for inverse reactions tend to zero. We introduce the
correspondent weak form of detailed balance, formulate the
necessary and sufficient algebraic conditions for this form of
detailed balance and formulate {the observables-initial data
symmetry for these systems.}

The experimental evidence of {the observables-initial data
symmetry} is presented for the reversible water gas shift
reaction over iron oxide catalyst. The experimental data are
obtained using Temporal Analysis of Products (TAP)
pulse-response studies. These offer excellent confirmation
within experimental error.

\section{Two forms of the reciprocal relations: forces, fluxes and entropic inner
product}

Let us consider linear kinetic equations or kinetic equations
linearized near an equilibrium $x^{\rm eq}$ (sometimes, it may
be convenient to move the origin to $x^{\rm eq}$):
\begin{equation}\label{kinur}
 \dot{x}=Kx  \, .
\end{equation}

In the original form of Onsager's relations, the vector of
fluxes $J$ and the vector of thermodynamic forces $X$ are
connected by a symmetric matrix, $J=LX$, $L_{ij}=L_{ji}$. The
vector $X$ is the gradient of the corresponding thermodynamic
potential: $X_i=\partial \Phi /\partial x_i$. For isolated
systems, $\Phi$ is the entropy. For other conditions, other
thermodynamic potentials are used. For example, for the
constant volume $V$ and temperature $T$ conditions, $\Phi$ is
$-F/T$ and for the constant pressure $P$ and temperature
conditions,  $\Phi$ is $-G/T$, where $F$ is the Helmholtz
energy (free energy) and $G$ is the Gibbs energy (free
enthalpy). These {\it free entropy} functions are also known as
the Massieu--Planck potentials \cite{Callen1985}. Usually, they
are {\em concave}.

For the finite-dimensional systems, like chemical kinetics or
the Master equation, the dynamics  satisfy linear (linearized)
kinetic equation $\dot{x}=K x$, where
$$K_{ij}=\sum_l L_{il}\left.\frac{\partial^2 \Phi}{\partial x_l
\partial x_j}\right|_{x^{\rm eq}} \mbox{ i.e. } K=L(D^2\Phi)_{x^{\rm eq}} .$$ This matrix is not
symmetric but the product $(D^2\Phi)_{x^{\rm eq}}K$
$=(D^2\Phi)_{x^{\rm eq}}L(D^2\Phi)_{x^{\rm eq}}$ is already
symmetric, hence, $K$ is symmetric (self-adjoint) in the
entropic scalar product
\begin{equation}\label{RecRel}
\langle a \, | \,K b\rangle_{\Phi}\equiv \langle K a \, | \,
b\rangle_{\Phi} \, ,
\end{equation}
where
\begin{equation}\label{thermcalprod}
\langle a \, | \, b\rangle_{\Phi}= -\sum_{ij}a_i \left.\frac{\partial^2 \Phi}{\partial x_l
\partial x_j}\right|_{x^{\rm eq}} b_j \, .
\end{equation}
Further on, we use the angular brackets for the entropic inner
product (\ref{thermcalprod}) and its generalizations and omit
the subscript $\Phi$.

For the spatially distributed systems with transport processes,
the variables $x_i$ are functions of the space coordinates
$\xi$, the equations of divergence form appear, $\partial_t
x_i= -\nabla_{\xi} \cdot J_i$, thermodynamic forces include
also gradients in space variables, $X_i=\nabla_{\xi}
\partial \Phi /\partial x_i$ and the operator $K$ has the form
$$
K_{ij}= \sum_l L_{il} \left.\frac{\partial^2 \Phi}{\partial
x_l
\partial x_j}\right|_{x^{\rm eq}}\Delta_{\xi} \, \mbox{ i.e. } K=L(D^2\Phi)_{x^{\rm eq}}\Delta_{\xi}  \, ,
$$
where $\Delta_{\xi}$ is the Laplace operator. This operator $K$
is self-adjoint in the inner product which is just the
integral in space of (\ref{thermcalprod}). The generalizations
to inhomogeneous equilibria, non-isotropic and non-euclidian
spaces are also routine but lead to more cumbersome formulas.

Symmetric operators have many important properties. Their
spectrum is real, for a function of a real variable $f$ with
real values it is possible to define $f(K)$ through the
spectral decomposition of $K$, and this $f(K)$ is also
symmetric in the same inner product. This property is the
cornerstone for further consideration.

\section{Symmetry between observables and initial data}

The exponential of a symmetric operator is also symmetric,
hence, Onsager's relations (\ref{RecRel}) immediately imply
\begin{equation}\label{ExpRecRel}
\langle a \, | \, \exp(Kt) \, b\rangle\equiv \langle b \, | \, \exp(Kt) \, a
\rangle \, .
\end{equation}
The expression $x(t)=\exp(Kt) \, b$ gives a solution to the
kinetic equations (\ref{kinur}) with initial conditions
$x(0)=b$. The expression $\langle a \, | \, x(t) \rangle$ is
the result of a measurement: formally, for each vector $a$ we
can introduce a ``device" (an observer), which measures the
scalar product of vector $a$ on a current state $x$.

The left hand side of (\ref{ExpRecRel}) represents the result
of such an experiment: we prepare an initial state $x(0)=b$,
start the process from this state and measure $\langle a \, |
\, x(t) \rangle$. In the right hand side, the initial condition
$b$ and the observer $a$ exchange their positions and roles: we
start from the initial condition $x(0)=a$ and measure $\langle
b \, | \, x(t) \rangle$. The result is the same function of
time $t$.

This exchange of the observer and the initial state transforms
an ideal experiment into another ideal experiment (we call them
{\it dual experiments}). The left and the right hand sides of
(\ref{ExpRecRel}) represent different experimental situations
but with the same results of the measurements.

This observation produces many consequences. As a first class
of examples, we present the time--reversible Markov chains
\cite{Hasegawa1976}, or the same class of kinetic equations,
the monomolecular reactions with detailed balance (see any
detailed textbook in chemical kinetics, for example,
\cite{Yablonskiiatal1991}).

Here a terminological comment is necessary. The term
``reversible" has three different senses in thermodynamics and
kinetics.
\begin{itemize}
\item{First of all, processes with entropy growth are {\em
    irreversible}. In this sense, all processes under
    consideration are irreversible.}
\item{Secondly, processes with microreversibility, which
    satisfy the detailed balance and Onsager relations, are
    {\em time--reversible} (or, for short, one often calls
    them ``reversible"). We always call them
    time--reversible to avoid confusion.}
\item{In the third sense, reversibility is the existence of
    inverse processes: if transition $A \to B$ exists then
    transition $B \to A$ exists too. This condition is
    significantly weaker than microreversibility.}
\end{itemize}
{ ``Time--reversibility" of irreversible processes sounds
paradoxical and requires comments. The most direct
interpretation of ``time--reversing" is to go back in time: we
take a solution to dynamic equations $x(t)$ and check whether
$x(-t)$ is also a solution. For the microscopic dynamics (the
Newton or Schr\"odinger equations) we expect that $x(-t)$ is
also a solution to the dynamic equations. Nonequilibrium
statistical physics combines this idea with the description of
macroscopic or mesoscopic kinetics by an ensemble of elementary
processes: collisions, reactions or jumps. The microscopic
``reversing of time" turns at this level into the ``reversing
of arrows": reaction $\sum_i \alpha_i A \to \sum_j \beta_j B_j$
transforms into $\sum_j \beta_j B_j \to \sum_i \alpha_i A$ and
conversely. The equilibrium ensemble should be invariant with
respect to this transformation. This leads us immediately to
the concept of {\em detailed balance}: each process is
equilibrated by its reverse process. ``Time--reversible kinetic
process" stands for ``irreversible process with the
time--reversible underlying microdynamics".}

We consider a general network of linear reactions. This network is
represented as a directed graph (digraph) \cite{Yablonskiiatal1991}:
vertices correspond to components $A_i$ ($i=1,2,\ldots,n$), edges
correspond to reactions $A_i \to A_j$ ($i\neq j$). For each vertex,
$A_i$, a positive real variable $c_i$ (concentration) is defined.
For each reaction, $A_i \to A_j$ a nonnegative continuous bounded
function, the reaction rate constant $k_{ji}> 0$ is given. The
kinetic equations have the standard Master equation form
\begin{equation}\label{kinur1}
\frac{\D c_i}{\D t}=\sum_{j, \, j\neq i} (k_{ij} c_j - k_{ji}c_i) \, .
\end{equation}
The principle of detailed balance (``time--reversibility")
means that there exists such a positive vector $c^{\rm eq}_i>0$
that for all $i,j$  ($j\neq i$)
\begin{equation}\label{lindetbal}
k_{ij}c^{\rm eq}_j=k_{ji}c^{\rm eq}_i \, .
\end{equation}
The following conditions are necessary and sufficient for
existence of such an equilibrium $c^{\rm eq}_i>0$:
\begin{itemize}
\item{Reversibility (in the third sense): if $k_{ji}> 0$
    then $k_{ij}> 0$;}
\item{For any cycle $A_{i_1}\to A_{i_2} \to \ldots \to
    A_{i_q} \to A_{i_1}$ the product of constants of
    reactions is equal to the product of constants of
    reverse reactions,
    \begin{equation}\label{CycleCond}
    \prod_{j=1}^q k_{i_{j+1} i_j}=\prod_{j=1}^q k_{i_j i_{j+1}} \, ,
    \end{equation}
    where $i_{q+1}=i_1$. {This is the {\em Wegscheider
    identity }\cite{Wegscheider1901}}.}
\end{itemize}
It is sufficient to consider in conditions (\ref{CycleCond}) a
finite number of basic cycles \cite{Yablonskiiatal1991}.

The free entropy function for the Master equation
(\ref{kinur1}) is the (minus) relative entropy
\begin{equation}\label{relative entropy}
Y=-\sum_i c_i \ln\left(\frac{c_i}{c^{\rm eq}_i}\right) \, .
\end{equation}
In this form, the function $-RTY$ was used already by L.~Onsager
\cite{Onsager1} under the name ``free energy". The entropic inner
product for the free entropy (\ref{relative entropy}) is
\begin{equation}\label{entrperfectproduct}
\langle a \, | \, b\rangle=\sum_i\frac{a_ib_i}{c^{\rm eq}_i} \, .
\end{equation}

Let $c^a(t)$ be a solution of kinetic equations (\ref{kinur1})
with initial conditions $c^a(0)=a$. Then the reciprocity
relations (\ref{ExpRecRel}) for linear systems with detailed
balance take the form
\begin{equation}\label{RecRelLinearBala}
\sum_i \frac{b_i c^a_i(t)}{c^{\rm eq}_i}=
\sum_i \frac{a_i c^b_i(t)}{c^{\rm eq}_i} \, .
\end{equation}
Let us use for $a$ and $b$ the vectors of the standard basis in
$\mathbb{R}^n$: $a_i=\delta_{iq}$, $b_i=\delta_{ir}$, $q\neq
r$. This choice results in the useful particular form of
(\ref{RecRelLinearBala}). We compare two experimental
situations, $c^a_i(0)= \delta_{iq}$ (the process starts at the
state ``everything is in $A_q$") and $c^b_i(0)= \delta_{ir}$
(the process starts at the state ``everything is in $A_r$");
for the first situation we measure $c^a_r(t)$ (``how much $A_r$
is produced from the initial $A_q$"), for the second one we
measure $c^b_q(t)$ (``how much $A_q$ is produced from the
initial $A_r$"). The reciprocal relations
(\ref{RecRelLinearBala}) give
\begin{equation}\label{RecRelLinearSimplest}
\frac{c^a_r(t)}{c^{\rm eq}_r}=\frac{c^b_q(t)}{c^{\rm eq}_q} \, .
\end{equation}
More examples of such relations for chemical kinetics are
presented in \cite{yab2010}. It is much more straightforward to
check experimentally these relations between kinetic curves
than the initial Onsager relations between kinetic
coefficients. We give an example of such an experiment below.
For processes distributed in space, instead of concentrations
of $A$ and $B$ some of their Fourier or wavelet coefficients
appear.

\section{Weak form of detailed balance}

For many real systems some of the elementary reactions are
practically irreversible. Hence the first condition of detailed
balance, the reversibility (if $k_{ji}> 0$ then $k_{ij}> 0$)
may be violated. Nevertheless, these systems may be considered
as {\em limits} of systems with detailed balance when some of
the constants tend to zero. For such limits, the condition
(\ref{CycleCond}) persists, and for any cycle the product of
constants of direct reactions is equal to the product of
constants of reverse reactions.

This is a {\em weak form of detailed balance} without the
obligatory existence of a positive equilibrium. In this
section, we consider the systems, which satisfy this weak
condition, the {\em weakly time--reversible} systems.

 For a linear system, the following condition is necessary and
sufficient for its weak time--reversibility:  {\em In any cycle
$A_{i_1}\to A_{i_2} \to \ldots \to A_{i_q} \to A_{i_1}$ with
strictly positive constants $k_{i_{j+1} i_j}>0$ (here
$i_{q+1}=i_1$) all the reactions are reversible ($k_{i_j
i_{j+1}}>0$) and the identity (\ref{CycleCond}) holds.}

The components $A_q$ and $A_r$ ($q\neq r$) are {\em strongly
connected} if there exist oriented paths both from $A_q$ to
$A_r$ and from $A_r$ to $A_q$ (each oriented edge corresponds
to a reaction with nonzero reaction rate constant). {It is
convenient to consider an empty path from $A_i$ to itself as an
oriented path.}

{\em For strongly connected components of a weakly
time--reversible system, all reactions in any directed path
between them are reversible.} This is a structural condition of
the weak time--reversibility.

Under this structural condition, the classes of strongly
connected components form a partition of the set of components:
these classes either coincide or do not intersect and each
component belongs to one of them. Each cycle belongs to one
class.

Let $A_q$ and $A_r$ be strongly connected. Let us select an
arbitrary oriented path $p$ between $A_q$ and $A_r$: $A_q
\leftrightarrow A_{i_1} \leftrightarrow A_{i_2} \leftrightarrow
\ldots \leftrightarrow A_{i_l} \leftrightarrow A_r$. For the
product of direct reaction rate constants in this path we use
$K_p^+$ and for the product of reverse reaction rate constants
we use $K_p^-$. {\em The ratio $K_{rq}=K_p^+/K_p^-$ does not
depend on the path $p$ and characterizes the pair $A_r,A_q$},
because of the Wegscheider identity (\ref{CycleCond}). This is
the quantitative criterion of the weak time--reversibility.

The constant $K_{rq}$ is an analogue to the equilibrium
constant. Indeed, for the systems with positive equilibrium and
detailed balance, $K_{rq}c^{\rm eq}_q=c^{\rm eq}_r$ and
$K_{rq}=c^{\rm eq}_r/c^{\rm eq}_q$.

For weakly time--reversible system, the reciprocal relations
between kinetic curves can be formulated for any strongly
connected pair $A_q$ and $A_r$. Exactly for the same pair of
kinetic curves,  as in (\ref{RecRelLinearSimplest}), we obtain
\begin{equation}\label{RecRelLinearSimplest2}
\frac{c^a_r(t)}{c^b_q(t)}=K_{rq} \, .
\end{equation}
This formula describes two experiments: (i) we start the system
at $t=0$ from the pure $A_q$ and measure $c_r(t)$, then (ii) we
start at $t=0$ from the pure $A_r$ and measure $c_q(t)$. The
ratio of these two kinetic curves, $c_r(t)/c_q(t)$ does not
depend on $t$ and is equal to the generalized equilibrium
constant $K_{rq}$.

{The weak form of the Wegscheider identity for general
(nonlinear) kinetic systems is also possible. Let us consider
the reaction system:
\begin{equation}\label{ReactMech}
\alpha_{r1} A_1+ \ldots +\alpha_{rn} A_n \to \beta_{r1} A_1+ \ldots
+\beta_{rn} A_n \, ,
\end{equation}
which satisfies the mass action law: $\dot{c}=\sum_r \gamma_r k_r
\prod_i c_i^{\alpha_i}\, ,$ where $k_r>0$,
$\gamma_{ri}=\beta_{ri}-\alpha_{ri}$ is the stoichiometric vector of
the $r$th reaction, and the reverse reactions with positive
constants are included in the list (\ref{ReactMech}) separately.

Let us consider linear relations between vectors $\{\gamma_r\}$:
\begin{equation}\label{RelationsGamma}
\sum_r \lambda_r \gamma_r =0 \mbox{ and } \lambda_r\neq 0 \mbox{ for
some } r\, .
\end{equation}
If all the reactions are reversible then the principle of
detailed balance gives us the identity
\cite{Yablonskiiatal1991}:
\begin{equation}\label{NonLinWeg}
\prod_r (k_r^+)^{\lambda_r}=\prod_r (k_r^-)^{\lambda_r}
\end{equation}
for any linear relation (\ref{RelationsGamma}). For reversible
reactions, we can take $\lambda_r\geq 0$ in (\ref{NonLinWeg}) for
all $r$: if we substitute the reactions with $\lambda_r < 0$ by
their reverse reactions then $\gamma_r$ and $\lambda_r$ change
signs. It is sufficient to consider only the cone $\Lambda_+$ of
non-negative relations (\ref{RelationsGamma}) ($\lambda_r\geq 0$)
and take in (\ref{NonLinWeg}) the direction vectors of its extreme
rays. Let $k_r^-=0$ for some $r$. The weak form of the identity
(\ref{NonLinWeg}) is:

{\em For any extreme ray of $\Lambda_+$ with a direction vector
$\lambda_r\geq 0$ the reactions which correspond to the positive
coefficients $\lambda_r > 0$ are reversible ($k_r^->0$) and their
constants satisfy the identity (\ref{NonLinWeg}).}

 }

{
\section{Nonlinear Examples}

It seems impossible to find a general relation between kinetic
curves for general nonlinear kinetics far from equilibrium.
Nevertheless, simple examples encourage us to look for a
nontrivial theory for some classes of nonlinear systems. In
this Section, we give two examples of nonlinear elementary
reactions which demonstrate the equilibrium relations between
nonequilibrium kinetic curves \cite{yab2010}.

\subsection{$2A\leftrightarrow B$}

The linear conservation law is $c_A+2c_B=const$. Let us take
two initial states with the same value $c_A+2c_B=1$: ($a$)
$c_A(0)=1 , \,c_B(0)=0$ and ($b$) $c_A(0)=0, \, c_B(0)=1/2$. We
will mark the corresponding solutions by the upper indexes
$a,b$. The mass action law gives:
\begin{equation}
\dot{c}_A=-2k^+c_A^2+k^-(1-c_A)\, , \; c_B=(1-c_A)/2 \, .
\end{equation}

 The analytic solution easily gives
\begin{equation}
\frac{c_B^a(t)}{c_A^a(t)
c_A^b(t)}=\frac{k^+}{k^-}=K^{\mathrm{eq}}=\frac{c_B^{\mathrm{eq}}}{(c_A^{\mathrm{eq}})^2}
\, ,
\end{equation}
the denominator involves the $A$ concentrations of both
trajectories, $c^a$ (started from $c_A(0)=1 , \,c_B(0)=0$) and $c^b$
(started from $c_A(0)=0, \, c_B(0)=1/2$). A ratio is equal to the
equilibrium constant at every time $t>0$. This identity between the
non-stationary kinetic curves reproduces the equilibrium ratio.

\subsection{$2A \leftrightarrow 2B$}

The linear conservation law is $c_A+c_B=const$. Let us take two
initial states with the same value $c_A+c_B=1$: ($a$) $c_A(0)=1
, \,c_B(0)=0$ and ($b$) $c_A(0)=0, \, c_B(0)=1$. The kinetic
equation is
\begin{equation}
\dot{c}_A=-2k^+c_A^2+k^-(1-c_A)^2\, , \; c_B={1-c_A} \, .
\end{equation}
It can be solved analytically. For this solution,
\begin{equation}
\frac{c_B^a(t) c_B^b(t)}{c_A^a(t) c_A^b(t)}
=\frac{k^+}{k^-}=K^{\mathrm{eq}}=\frac{(c_B^{\mathrm{eq}})^2}{(c_A^{\mathrm{eq}})^2}
\, ,
\end{equation}
both the numerator and denominator include trajectories for both
initial states, $a$ and $b$. This identity between the kinetic
curves also reproduces the equilibrium ratio.

}

\section{Experimental evidences}

In this work, we investigate the validity of the reciprocal
relations using the TAP (Temporal Analysis of Products)
technique proposed by Gleaves in 1988 \cite{yab1988}. It has
been successfully applied in many  areas of chemical kinetics
and engineering for non-steady-state kinetic characterization
\cite{yab2003}. The studied reaction  is a part of  the
reversible water gas shift reaction over iron oxide catalyst.
The overall reaction is $\mbox{H}_2\mbox{O}+\mbox{CO}
\leftrightarrow \mbox{H}_2+\mbox{CO}_2$.

\subsection{Experimental set-up}

The TAP reactor system used in this work is made of quartz and
is of the size 33 mm bed-length and 4.75 mm inner diameter. The
products and the unreacted reactants coming out of the reactor
are monitored by a UTI 100C quadrupole mass spectrometer (QMS).
The number of molecules admitted during pulse experiments
amounts to $10^{15}$ molecules/pulse.

To ensure uniformity of the catalyst along the bed, we use a
thin--zone TAP reactor (TZTR), the width of the catalyst zone
being 2mm. Experiments were performed over 40 mg of
Fe${}_2$O${}_3$ catalyst. The catalyst was packed in between
two inert zones of quartz particles of the same size
($250<d_p<500\mu m$). The temperature of the reactor was
measured by a thermocouple positioned in the center of the
catalyst bed. Several single pulse experiments were performed
by pulsing CO or CO${}_2$ at the temperature of 780K. In all
the experiments, the reaction mixture was prepared with Ar as
one of the components, so that the inlet amount of the
components can be determined from the Ar response.

\subsection{Application to the measurements}

\begin{figure}[t]
\begin{center}
\includegraphics[width=0.47\textwidth]{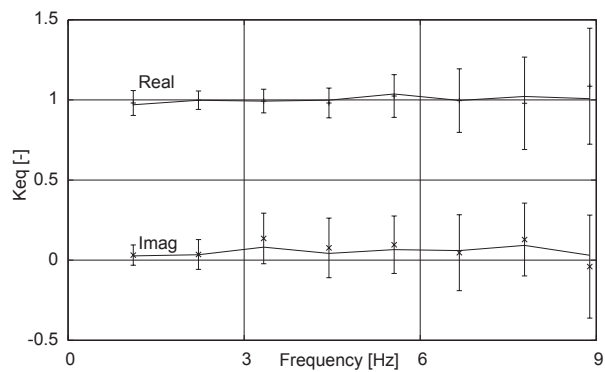}
\end{center}
\caption{\label{fourfig}Fourier domain result values for the
 ``B from A/A from B" ratio (\ref{motherlode}), vs.\ frequency $f$ in Hz (so that $
\omega=2\pi f$, $s=i\omega$);  real and imaginary part. The
error bars were obtained from 10,000 resampled measurements.}
\end{figure}

In a thin--zone TAP-reactor, the diffusion occurring in the
inert zones flanking the thin reactive zone must be accounted
for. The Knudsen regime in these zones guarantees a linear
behaviour, so that the resulting outlet fluxes can be expressed
in terms of convolutions. Switching to the Laplace domain
greatly facilitates the analysis, and we can prove in general
that the fixed proportion property is equivalent to the
following equality in terms of the exit fluxes $F_{B_A}$ of gas
$B$ given a unit inlet pulse of gas $A$ and $F_{A_B}$, of $A$
given a unit inlet pulse of $B$, see \cite{constales2004}:
\begin{equation}
K_{\mathop{\rm eq}}=
{\frac{(\cosh\sqrt{s\tau_{1,A}})(\sqrt{\tau_{3,B}}\sinh\sqrt{s\tau_{3,B}})}
{ (\cosh\sqrt{s\tau_{1,B}})(\sqrt{\tau_{3,A}}\sinh\sqrt{s\tau_{3,A}})}}
{\frac{{\cal L}F_{B_A}(s)}{{\cal L}F_{A_B}(s)}}
\label{motherlode}
\end{equation}
identically in the Laplace variable $s$, where  $\tau_{i,G}=
\epsilon_iL_i^2/D_G$,  $\epsilon_i$ denoting the packing
density of the $i$-th zone, $L_i$ its length, and $D_G$ the
diffusivity of gas $G$. To apply this in practice, we set
$s=i\omega$ and switch to the Fourier domain.

Performing these corrections, with $A$ denoting CO and $B$ CO$
{}_2$, the results of Fig.~\ref{fourfig} in the Fourier domain
are obtained. The real and imaginary parts of the right-hand
side in (\ref{motherlode}) are graphed, with error bars
corresponding to three times the standard deviation estimated
from resampling 10,000 times the exit flux measurements using
their principal error components. Ideally, all imaginary values
should be zero; we see that zero does lie within all the
confidence intervals. We also see that the sma\-l\-lest error
in the real parts occurs for the second fre\-quency, 2.2 Hz.
This confidence interval lies snugly within the others,
offering confirmation that (within experimental error) the same
value for all frequencies is obtained.

\section{Conclusion}

The shift in time operator is symmetric in the entropic inner
product. Its symmetry allows us to formulate the symmetry relations
between the observables and initial data. These relations could be
validated without differentiation of empiric curves and are, in that
sense, more robust and closer to the direct measurements. For the
Markov processes and chemical kinetics, the symmetry relations
between the observables and initial data have an elegant form of the
symmetry between ``$A$ produced from $B$" and ``$B$ produced from
$A$": their ratio is equal to the equilibrium constant and does not
change in time (\ref{RecRelLinearSimplest}),
(\ref{RecRelLinearSimplest2}). For processes distributed in space,
instead of concentrations of $A$ and $B$ some of their Fourier or
wavelet coefficients appear.

The symmetry relations between the observables and initial data have
a rich variety of realizations, which makes the direct experimental
verification possible. On the other hand, this symmetry provides the
possibility to extract information about the experimental data
through the dual experiments. These relations are applicable to all
systems with microreversibility.

\end{document}